\documentclass[dvips]{article}

\usepackage{icrc2011}

\title{UHECR by lightest nuclei in Nearby Universe and its parasite neutrino trace}
\shorttitle{D. Fargion,  UHECR and Neutrino Astronomy}

\authors{Daniele Fargion$^{1,2}$ }
\afiliations{$^1$Physics Department, Rome University 1, Sapienza,  Ple.A.Moro 2, Rome,Italy\\ $^2$INFN Rome 1 }
\email{daniele.fargion@Roma1.infn.it}

\abstract{UHECR mystery survived first Auger claims of AGN connection within a GZK (100 Mpc
size) Universe. Last 2010 UHECR maps and compositions do show a  much lower
correlation with AGN SuperGalactic maps and with protons (the two main ingredient of
that claim). The three main AUGER results that survived are: an embarrassing absence of
Virgo cluster UHECR, an expected signal;  a steady presence of Cen-A clustering and a
remarkable signature of nuclei (not nucleon) composition. We claim that the He-like
lightest nuclei may solve most of the puzzle: He UHECR cannot arrive from Virgo
because light nuclei fragility and opacity; Cen-A UHECR are spread (if He-Li-Be) as
much as the observed ones; the light nuclei may fit well Auger composition signature as
well as Hires spectra. Future multiplet of half energy UHECR must trace in tails
some of UHECRa. Their secondaries GZK neutrinos will crowd not at thousand PeV, but
more at tens PeV observable maybe in ICECUBE and also in horizontal tau airshowers in ARGO or by upward tau airshowers in
Auger and TA fluorescence telescopes. }
\keywords{ The keywords will be used to select your subject from all
ICRC contributions. }

\begin{document}
\maketitle

%Begin the section.

\section{UHECR by He and the fragments}
UHECR astronomy is becoming a reality, with some confusion because of the magnetic field smearing of their arrival directions. Moreover
  while flying UHECR are making photo-pion (if nucleon) or gamma and neutrinos by photo-dissociation (if light nuclei). Making UHECR nucleon local and sharp (GZK cut off, tens Mpc) or very local and smeared (a few Mpc) for lightest nuclei, or making UHECR astronomy meaningless if made by iron or other heavy nuclei. Therefore any  UHECR astronomy is surrounded by a parasite astronomy made by gamma and neutrinos as well as by their possible small radio tails and  UHECR fragments. Indeed UHECR formed by  lightest nuclei may explain clustering of events around CenA and a puzzling UHECR absence  around Virgo. Their fragments $He + \gamma \rightarrow D+D, He + \gamma \rightarrow He^{3}+n, He + \gamma \rightarrow T +p $  may trace on the same UHECR maps by a secondary tail or a crown clustering at half or fourth the UHECR primary  energy. Neutrinos and gamma are tracing  (both for nucleon or nuclei) their UHECR trajectory, respectively at EeVs or PeV energy. Gamma rays are partially absorbed by microwave and infrared background making only a very local limited astronomy. Among neutrinos $\nu$,  muons ones $\nu_{\mu}$, the most penetrating and easy to detect, are deeply polluted by atmospheric  component (smeared and isotropic like their parent CR ). As shown by last TeV muon neutrino maps  probed by very smooth ICECUBE neutrinos map. Tau neutrinos,  the last neutral lepton discovered, absent in neutrino oscillation at TeVs-PeVs-EeVs atmospheric windows,  may arise as the first clean signal in UHECR-neutrino associated astronomy. Their tau birth in ice  may shine as a double bangs (disentangled above PeV) anisotropy. In addition UHE tau, born tangent to the  Earth or mountain, while escaping  in air  may lead, by decay in flight, to  loud, amplified  well detectable tau-airshower at horizons. Tau astronomy versus UHECR are going to reveal most violent  sky as the most deepest probe. First hint of  Vela, the brightest and nearest gamma source, a first galactic source  is rising as a UHECR triplet nearby. Cen A (the most active and nearby AGN) is apparently shining  UHECR source whose clustering (almost a quarter of the event) along a narrow solid angle around (whose opening angular size is  $\simeq 17^{o}$) seem  firm and it is  favoring lightest nuclei. Remaining  events are possibly more  smeared being more bent and heavier nuclei  of  galactic and-or  extragalactic origin.
The rise of nucleon UHECR above GZK astronomy made by protons (AUGER November 2007) is puzzled by three main mysteries: an unexpected nearby Virgo UHECR suppression (or absence), a rich crowded clustering frozen vertically along Cen A, a composition suggesting nuclei (not much directional) and not nucleons. The UHECR map, initially consistent with  GZK volumes, to day seem to be not much correlated with expected Super Galactic Plane. Moreover slant depth data of UHECR from AUGER airshower shape do not favor the proton but points to  a nuclei. To make even more confusion (or fun) HIRES, on the contrary, seem to favor, but with less statistical weight, UHECR mostly nucleons. We tried  (at least partially) to solve the contradictions assuming UHECR as light nuclei ( $He^4$, Li, Be)) spread by planar galactic fields, randomly at vertical axis. The $He^4$ fragility and its mass and its charge explains naturally the Virgo absence (due to $He^4$ opacity above few Mpc) and the observed wide Cen A spread clustering  (a quarter of the whole sample within $17^{o}$). However more events and rarest doublets and  clustering are waiting for an answer. Here we foresee hint of a new UHECR component, due to a first timid triplet toward Vela, the nearest and the brightest gamma source (at few or tens GeV) as well as the possible first high energy source (cosmic ray at tens TeV). Indeed early anisotropy of tens TeVs Cosmic rays found in ICECUBE muons might confirm this possibility.
Let us remind that last century have seen the birth of a puzzling cosmic ray whose nature and origination has been and it is still growing in an apparent never-ending chain of  puzzle.
The Cosmic Black Body Radiation had imposed since $1966$ a cut, GZK cut-off \cite{Greisen:1966jv}, of highest energy cosmic ray propagation, implying a very limited cosmic Volume (ten or few tens Mpc) for highest UHECR  nucleon events. Because of the UHECR rigidity one had finally to expect to track easily UHECR directionality back toward the sources into a new Cosmic Rays Astronomy.
Indeed in last two decades, namely since 1991-1995 the rise of an \emph{apparent} UHECR at $3 {10^{20}}$ eV, by Fly's Eye,  has opened the wondering of its origination: no nearby (within GZK cut off) source have been correlated. Incidentally it should be noted that even after two decades  and after an increase of aperture observation  by nearly two order of magnitude (area-time by AGASA-HIRES-AUGER) no larger or equal event as $3 {10^{20}}$ eV has been rediscovered. Making wondering the nature of that exceptional starting UHECR event. To face the uncorrelated UHECR at $3 {10^{20}}$ eV, and later on event by AGASA, the earliest evidences (by SuperKamiokande) that neutrino have a non zero mass had opened  \cite{Fargion1997} the possibility of  an UHECR-Neutrino connection:  UHE ZeV neutrino could be the transparent  courier of a far AGN (beyond GZK radius) that may hit and scatter on  a local relic anti-neutrino dark halo, spread  at a few Mpc around our galaxy. Its resonant Z boson (or WW channel) production  \cite{Fargion1997} is source of a secondary nucleon later on observable at Earth as a UHECR.  The later search by AGASA seemed to confirm the Fly's Eye by events above $ {10^{20}}$eV  and the puzzling absence of nearby expected GZK anisotropy or correlation.  On $1999-2000$ we were all convinced on the UHECR GZK cut absence. In different occasion HIRES data offered a possible  UHECR connections with far BL Lac \cite{Gorbunov}, giving argument to such Z-resonant (Z-burst) model. However more recent  records by a larger Hires area  (2001-2005)  have been claiming  evidences of GZK suppression in UHECR spectra. The same result seemed confirmed by last AUGER data in last few years \cite{Auger-Nov07}. But in addition AUGER have shown an anisotropic clustering, seeming along the Super Galactic Plane, a place well consistent with GZK expectation \cite{Auger-Nov07}. Because of it the UHE neutrino scattering model \cite{Fargion1997},) became obsolete.

 Indeed here we review the most recent  UHECR maps \cite{Auger10} over different Universe   and we comment some  feature, noting some possible minor galactic component \cite{Fargion09b}, \cite{Fargion2010}. Moreover we remind possible Z-Showering model solution if UHECR are correlated to AGN,BLac or Quasars at large redshift. The consequences of the UHECR composition and source reflects into UHE (GZK \cite{Greisen:1966jv} or cosmo-genic) neutrinos. The proton UHECR provide EeV neutrinos (muons and electron) whose flavor oscillation lead to tau neutrinos to be soon detectable \cite{FarTau} \cite{Auger08} by upward tau air-showers;  the UHECR lightest nuclei model provide only lower energy, tens PeV, neutrinos detectable in a very peculiar way by AUGER fluorescence telescopes or in ARGO array by horizontal $\tau$ air-showers  , or by Icecube $km^3$ neutrino telescopes \cite{FarTau},\cite{Fargion2009},\cite{Fargion09a}  \cite{Fargion09b} either by double bang\cite{Learned}, or by long muon at few PeV energy. ZeV UHE neutrinos in Z-Shower model are possible source of horizontal Tau air-showers of maximal size and energy \cite{FarTau},\cite{Fargion2010}.
\subsection{The Lorentz UHECR bending and spread}
%%%%%%%%%%%%%%%%%%%%%%%%%%%%%%%%%%%%%%%%%%%%%%%%%%%%%
Cosmic Rays are blurred by magnetic fields. Also UHECR suffer of a Lorentz force deviation. This smearing maybe source of UHECR features. Mostly along Cen A. There are two main spectroscopy of UHECR along galactic plane:
 A late nearby (almost local) bending by a nearest coherent galactic arm field, and a random one along the whole plane inside different arms.
The coherent Lorentz angle bending $\delta_{Coh} $ of a proton UHECR (above GZK \cite{Greisen:1966jv}) within a galactic magnetic field  in a final nearby coherent length  of $l_c = 1\cdot kpc$ is $ \delta_{Coh-p} \simeq{2.3^\circ}\cdot \frac{Z}{Z_{H}} \cdot (\frac{6\cdot10^{19}eV}{E_{CR}})(\frac{B}{3\cdot \mu G}){\frac{l_c}{kpc}}$.
The corresponding coherent  bending of an Helium UHECR at same energy, within a galactic magnetic field
  in a wider nearby coherent length  of $l_c = 2\cdot  kpc$ is
\begin{equation}
\delta_{Coh-He} \simeq
{9.2^\circ}\cdot \frac{Z}{Z_{He}} \cdot (\frac{6\cdot10^{19}eV}{E_{CR}})(\frac{B}{3\cdot \mu G}){\frac{l_c}{2 kpc}}
\end{equation}

%%%%%%%%%%%%%%%%%%%%%%%%%%%%%%%%%%%%%%%%%%%%%%%%%%%%%%%%%%%%%%%

This bending angle is compatible with observed multiplet along $Cen_A$ and also the possible clustering along Vela, at much nearer distances; indeed in latter case it is possible for a larger magnetic field along its direction (20 $\mu G$) and-or for a rare iron composition $\delta_{Coh-Fe-Vela} \simeq {17.4^\circ}\cdot \frac{Z}{Z_{Fe}} \cdot (\frac{6\cdot10^{19}eV}{E_{CR}})(\frac{B}{3\cdot \mu G}){\frac{l_c}{290 pc}}$. Such iron UHECR are mostly bounded inside a Galaxy, as well as in Virgo, explaining partially  its extragalactic absence. In lightest nuclei model the rare heavier iron nuclei  may be bounded inside  Virgo.  The incoherent random angle bending, $\delta_{rm} $, while crossing along the whole Galactic disk $ L\simeq{20 kpc}$  in different spiral arms  and within a characteristic  coherent length  $ l_c \simeq{2 kpc}$ for He nuclei is $\delta_{rm-He} \simeq{16^\circ}\cdot \frac{Z}{Z_{He^2}} \cdot (\frac{6\cdot10^{19}eV}{E_{CR}})(\frac{B}{3\cdot \mu G})\sqrt{\frac{L}{20 kpc}} \sqrt{\frac{l_c}{2 kpc}}$ The heavier  (but still lightest nuclei) bounded from Virgo are Li and Be:
$\delta_{rm-Li} \simeq {24^\circ}\cdot \frac{Z}{Z_{Li^3}} \cdot (\frac{6\cdot10^{19}eV}{E_{CR}})(\frac{B}{3\cdot \mu G})\sqrt{\frac{L}{20 kpc}}
\sqrt{\frac{l_c}{2 kpc}} $, $\delta_{rm-Be} \simeq{32^\circ}\cdot \frac{Z}{Z_{Be^4}} \cdot (\frac{6\cdot10^{19}eV}{E_{CR}})(\frac{B}{3\cdot \mu G})\sqrt{\frac{L}{20 kpc}}
\sqrt{\frac{l_c}{2 kpc}}$.  It should be noted that the present anisotropy above GZK \cite{Greisen:1966jv} energy $5.5 \cdot 10^{19} eV$ might leave a tail of signals: indeed the photo disruption of He into deuterium, Tritium, $He^3$ and protons (and unstable neutrons), might rise as clustered events at half or a fourth of the energy:\emph{ being with a fourth an energy but half a charge proton tails will smear around Cen-A at twice larger angle}. It is important to look for correlated tails of events, possibly in  strings at low $\simeq 1.5-3 \cdot 10^{19} eV$ along the $Cen_A$ train of events. \emph{It should be noticed that Deuterium fragments are half energy and mass of Helium: Therefore D and He spot are bent at same way and overlap into UHECR circle clusters}.  Deuterium are even more bounded in a local Universe because their fragility. In conclusion He like UHECR  maybe bent by a characteristic as large as  $\delta_{rm-He}  \simeq 16^\circ$,( its expected lower energy proton tails at $\delta_{rm-p}  \simeq 32^\circ$). Well within the observed CenA UHECR clustering spread.

 \begin{figure}[!t]
  \vspace{5mm}
  \centering
  \includegraphics[width=2.2 in]{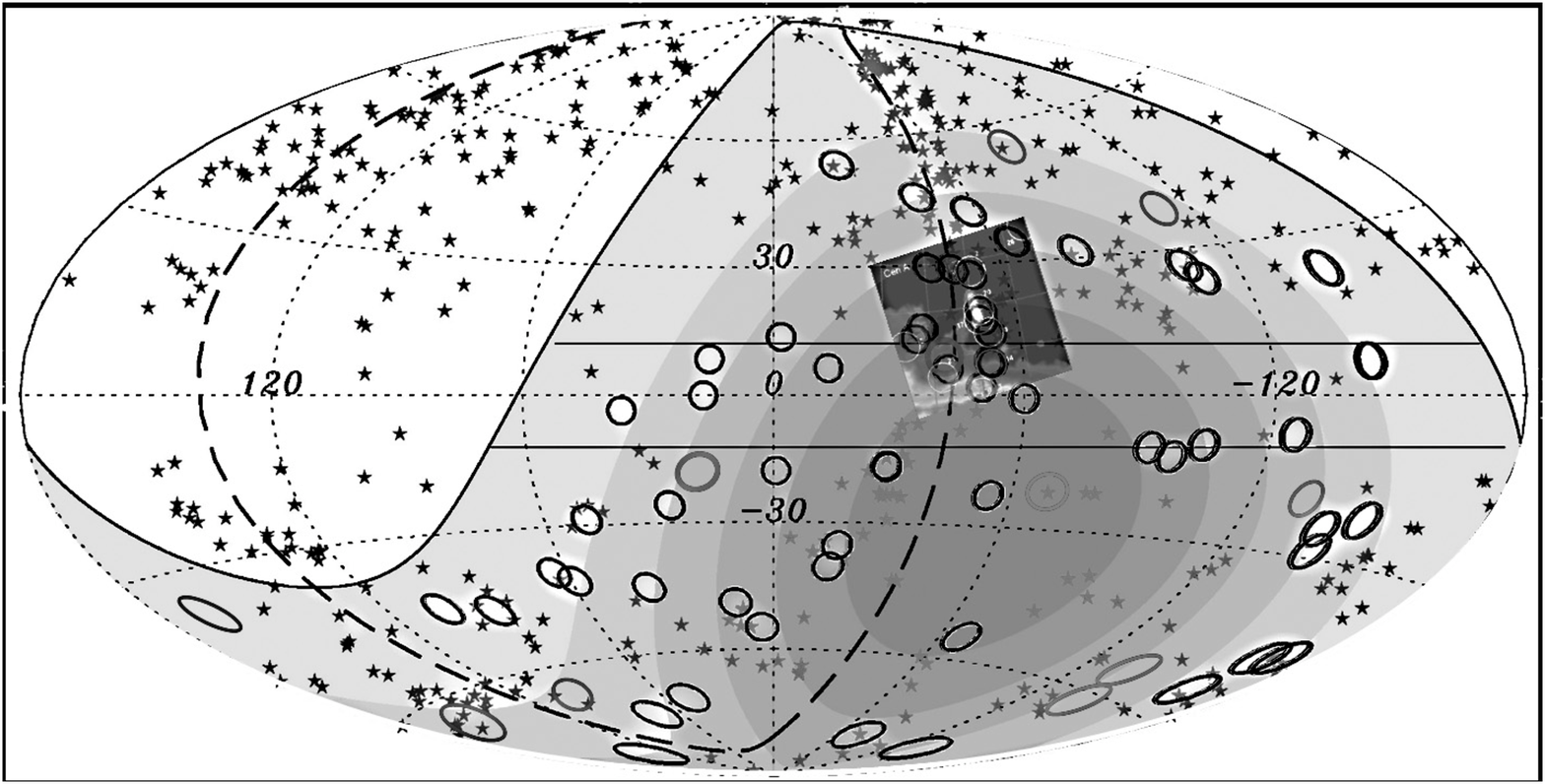}
  \caption{The last 2010 UHECR event map by AUGER and the clustering toward Cen A, nearest AGN-Jet}
  \label{simp_fig}
 \end{figure}

 \begin{figure}[!t]
  \vspace{5mm}
  \centering
  \includegraphics[width=2.3 in]{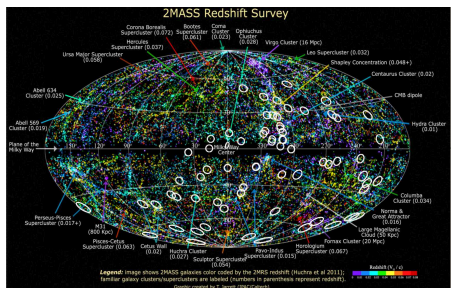}
  \caption{The last 2010 UHECR event map by AUGER and the most recent Two Micron All Sky Survey (2MASS); it is remarkable the Virgo absence also in view of most recent TA and HIRES data.
  The cosmic  Universe inside GZK volume do not correlate much with UHECR out of nearest Cen-A  AGN and partial galactic possible components. }
  \label{simp_fig}
 \end{figure}

 \begin{figure}[!t]
  \vspace{5mm}
  \centering
  \includegraphics[width=2.2 in]{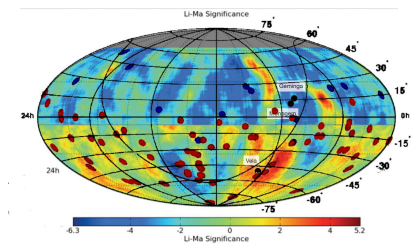}
  \caption{A first hint of  Vela as a UHECR  in celestial coordinated . The 69 Auger events (red-gray disks) are also shown with the 13 Hires (blue-black disk) UHECR events; note Virgo absence, note the clustering along Cen-A. Note the TeVs CR clustering around Vela, the brightest Gamma source in our galaxy with a triplet UHECR along.}
  \label{simp_fig}
 \end{figure}

 \begin{figure}[!t]
  \vspace{5mm}
  \centering
  \includegraphics[width=2.2 in]{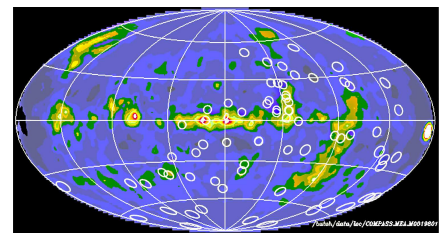}
  \caption{The gamma Osse (Compton Observatory) map (1-3 MeV energy band)  with the 69 UHECR of AUGER. This MeV background radiation is mostly of galactic sources as isotopes of titanium (Ti-44) and aluminum (Al-26), which are both produced in supernova explosions. Other sources are (AGN), which are  massive black holes  mostly (excluding CenA) at distances much larger than GZK. The clustering around CenA and the triplet around Vela, are correlated both in Comptel map; Vela triplet is correlated with new 12 TeV muon anisotropy discovered by IceCube \cite{Desiati}.}
  \label{simp_fig}
 \end{figure}

\begin{figure}[!t]
  \vspace{5mm}
  \centering
  \includegraphics[width=2.1 in]{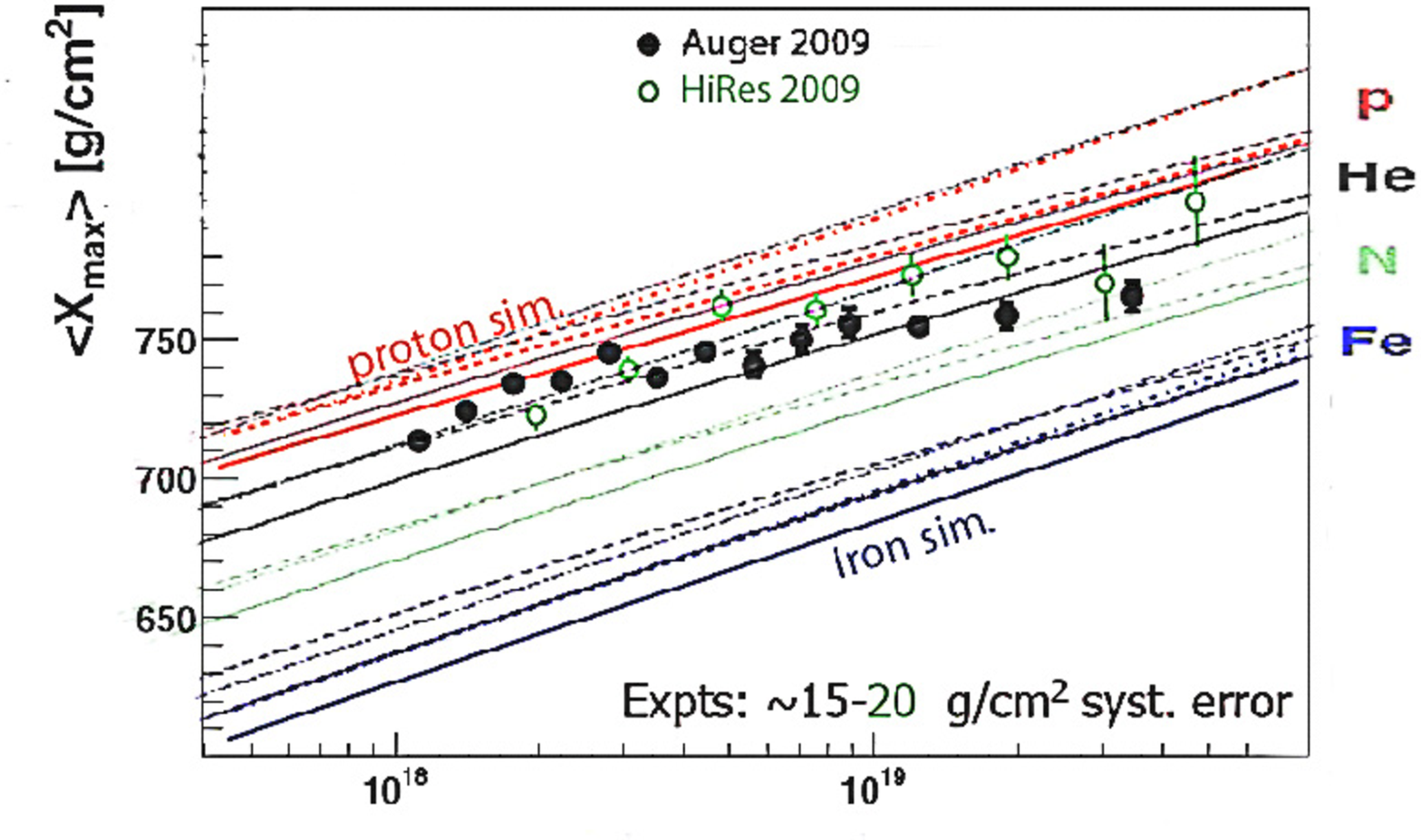}
  \caption{The last UHECR AUGER Composition derived by air shower slant depth ; note the best fit of He on most of the highest UHECR events}
  \label{simp_fig}
 \end{figure}

 \begin{figure}[!t]
  \vspace{5mm}
  \centering
  \includegraphics[width=2.1 in]{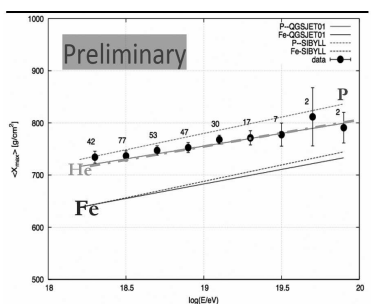}
  \caption{The last UHECR Telescope Array Composition derived by air shower slant depth shown in RICAP 2011; note the best fit of He on most of the highest UHECR events}
  \label{simp_fig}
 \end{figure}

 \begin{figure}[!t]
  \vspace{5mm}
  \centering
  \includegraphics[width=2.4 in]{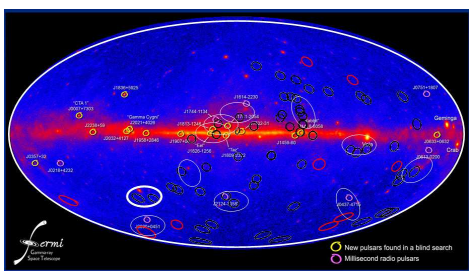}
  \caption{  The very recent Flare from far AGN blazar $3C454.3$, at half the Universe distance in Fermi sky and with UHECR. Its relevance is not just related to the huge output of the source and to the doublet AUGER event connected by this map: but also to additional signals. The Z-resonant (or Z-Burst) model explains this otherwise mysterious connection. The UHE neutrino primary energy need to be nearly $10-30$ ZeV and the relic neutrino mass might range in the $0.4-0.133$ eV. The whole conversion efficiency might range from a minimal $10^{-4}$ \cite{Fargion1997} for no relic neutrino clustering up to $4 \cdot 10^{-3}$ a (forty) times density contrast in Local Group halo. Even within the minimal conversion efficiency, observed  gamma flaring $3C454.3$ blazar  is consistent with the extreme UHECR flux assuming a primary Fermi flat spectra (of the blazar) extending up to ZeV energy. See \cite{Fargion1997}.Other AGN or Vela correlation are drawn by oval rings.}
  \label{simp_fig}
 \end{figure}

\section{Conclusions: UHECR and UHE neutrinos}
 The history of Cosmic Rays and last UHECR discoveries (and disclaims) are exciting and surprising. The very surprising correlation with Cen A, the absence of Virgo, the hint of correlation with  Vela and galactic center might be solved by a lightest nuclei, mainly He, as a  courier, leading to a very narrow (few Mpc) sky for UHECR. However the very  exceptional  blazar $3C454.3$ flare on $2nd December$ $2009$, the few AGN connection of UHECR located much far from a GZK distances may force us, surprisingly, to reconsider an exceptional model: Z-Shower one \cite{Fargion1997}. Possibly connecting lowest neutrino   particle ($\simeq 0.15$ eV) mass with highest UHE ($\simeq 30$ZeV)neutrino energies. Even for the minimal UHE $\nu$-Z-UHECR conversion as low as $10^{-4}$ (see table$1$,last reference in\cite{Fargion1997}), \emph{for a not clustered relic neutrino halo as diluted as cosmic ones}  the present gamma $3C454.3$  output (above $3\cdot 10^{48}$ $erg s^{-1}$) is  comparable with UHECR (two events in $4 years$ in AUGER), assuming  a flat Fermi spectra (for neutrinos) extended up to UHECR ZeVs edges. These results are somehow surprising and revolutionary. We might be warned for unexpected very local and a very wide  Universe sources sending UHECR traces in different ways. Testing finally the most evanescent \emph{hot} neutrino relic background. The consequences may be soon detectable in different way by Tau air-showers in AUGER,TA, Heat and also in unexpected horizontal shower in deep valley by ARGO \cite{FarTau},\cite{Auger08} or in widest atmosphere layer on Earth, Jupiter  and Saturn, see papers in\cite{Fargion1997}. A soon answer maybe already written into present  clustering (as Deuterium fragments) at half UHECR edge energy around or along main UHECR group seed. Indeed He like UHECR  maybe bent by a characteristic as large as  $\delta_{rm-He}  \simeq 16^\circ$,(while expected  proton at half fourth these energies, along tails spread  at $\delta_{rm-p}  \simeq 32^\circ$).  The next release of UHECR update events, possibly above 2-3 tens EeV, their fragment clustering maps along higher energy UHECR (5-6 $10^{19}$ eV) may solve the puzzles. As well the eventual revolutionary discover of an horizontal  tau airshower nearby (by tau neutrinos at PeVs by UHECR light nuclei photo-dissociations) observed by ARGO; or by near (PeVs) or far (EeV) tau airshowers observed from fluorescence Auger and TA telescopes \cite{Fargion09b}.

\clearpage


\begin{thebibliography}{99}
\bibitem{Auger-Nov07} Pierre Auger Collaboration, \emph{Science}\textbf{vol.318},  939-943,(2007);
\bibitem{Auger08} Auger Collaboration,  \emph{Phys. Rev. Letters} \textbf{100}211101, \textbf{arXiv:0903.3385v1},(2009)
\bibitem{Auger10} Auger Collaboration,  \emph{Astroparticle Physics on 31 August 2010}, \textbf{arXiv:1009.1855v2}
\bibitem{Auger-Nov09} J.~ Cronin for AUGER, (2009) \textbf{arXiv:0911.4714}
\bibitem{Local09} A. J. Cuesta. and F. Prada, \textbf{arXiv:0910.2702v1}
\bibitem{Fargion1997}  D.~Fargion, B.Mele, A.Salis;  astro-ph/9710029, \emph{Astrophys.J.} \textbf{517}725-733 (1999);
D.Fargion, et al.\emph{Nucl.Phys. B}, \textbf{168}; 292-295;\emph{Nucl.Phys. B}, \textbf{165}(2007) 116,121,(2007); D.~Fargion \emph{Phys. Soc. Jpn.}\textbf{77}:92-100,(2008); D.~Fargion et al.  \emph{Phys. Soc. Jpn.}\textbf{70} 46-57 (2001);  D.~Fargion et al.  \emph{Recent Res.Devel.Astrophys.} \textbf{1},1(2003): 395-454:
\bibitem{FarTau}  D.~Fargion,\textbf{astro-ph/9704205v1},\emph{ApJ}, \textbf{570}, 909 (2002); D. Fargion~, et all. \emph{ApJ}, \textbf{613}, 1285,(2004);
\bibitem{Fargion2008}D.~ Fargion  \emph{Phys. Scr.} \textbf{78},  045901, 1-4.(2008).
\bibitem{Fargion09a}  D.~Fargion et al., \emph{Nuclear Physics B} \textbf{190},162 (2009).
\bibitem{Fargion09b}  D.~Fargion,  \emph{ Progress in Particle and Nuclear Physics 64 (2010) 363-365}
\bibitem{Fargion2009}  D.~Fargion, D. D'Armiento, P. Paggi, S. Patri' \emph{Nuclear Physics B (Proc. Suppl).} \textbf{190} 162-166(2009)
\bibitem{Fargion2010} 	D.~Fargion, \emph{AIP Conf. Proc.} March 26,Volume \textbf{1223}, pp. 149-158 (2010), arXiv:1001.1547v1.
\bibitem{Gorbunov}  D~ S~ Gorbunov, et al.  \emph{Astrophys.J.L} \textbf{577} L93( 2002)
\bibitem{Gorbunov09} D. Gorbunov, P.Tinyakov, I. Tkachev, S. Troitsky, \emph{JETP Lett.}\textbf{87}:461-463,(2008)
\bibitem{Greisen:1966jv} K. Greisen~ 1966  \emph{Phys. Rev. Lett.} \textbf{16} 748,  G.T. Zatsepin,, V.A. Kuz'min, \emph{Zh. Eks. Teor.Fiz., Pis'ma Red.} \textbf{4}144  (1966)
\bibitem{Learned} J G  Learned, Pakvasa S \emph{Astropart. Phys.}\textbf{3}, 267,(1995)
\bibitem{Desiati}  Abbasi R., Desiati P.,Astrophys.J.718:L194,2010
\bibitem{Semikoz10}   D. Semikoz \textbf{arXiv:1010.2647}; 

%%%%\bibitem{Yoshida1998}  Yoshida, S.,et all,\emph{Phys. Rev. Lett.} \textbf{81},5505-5508(1998).

%%%%%%%%%%%%%%%%%%%%%%%%%%%%%%%%%%%%%%%%%%%%%%%%%%%%%%%%%%%%%%%%%%%%%%%%%%%%%%%%%%%%%%%%%5

%%%%%%%\bibitem{Fargion1999}{\normalsize D.Fargion,et al.26th ICRC,He 6.1.09,p.396-398.1999.(USA);}
%%%\bibitem{Fargion2000}{\normalsize  D.Fargion~, 2002, ApJ, 570, 909;  D. Fargion~, et all. 2004, ApJ, 613, 1285;~D.Fargion et al., Nuclear Physics B %%%%(Proc. Suppl.)2004;~ D. Fargion, et all. Adv. in Space Res.,37 (2006) 2132-2138;,136 ,119; D.Fargion, J.Phys.Soc.Jpn.Vol.77 (2008) Suppl. B., %%%%%p.1-15. D.Fargion~ Prog. Part. Nucl. Phys 57,2006,384-393;~ D. Fargion et al. Adv.Space Res. 37 (2006) 2132-2138;~ D.Fargion et al. Nucl.Inst. %%%%%%Meth.A, 588; 2008, 146-150.}
%%%\bibitem{Fargion2009}{\normalsize  D.Fargion, D. D'Armiento,et al., Nuclear Physics B (Proc. Suppl.) 190 (2009) 162-166}
%%%%%\bibitem{Fargion2009b}{\normalsize  D.Fargion, D. D'Armiento,Venice "Neutrino Telescopes" 09,arXiv:0905.1517 }
%%%%%\bibitem{Feng2002}{\normalsize J. L. Feng1, P. Fisher, F. Wilczek,T.M. Yu,Phys. Rev. Lett. 88, 161102 (2002)}

%%%%\bibitem{Bertou2002}{\normalsize ~ X. Bertou et.all 2002, Astropart. Phys., 17,183}
%%%%%\bibitem{Bigas07}{\normalsize B. Blanch, Auger Coll.,ICRC07,arXiv:0706.1658v1}
%%%%\bibitem{Hires06}{\normalsize The HIRES Collaboration, Ap.J. 636, 2006, 680-684}
%%%%%\bibitem{Medina07}{\normalsize Medina Tanco G. for Auger Coll.  arXiv0709.0772M}
%%%%\bibitem{za66}{\normalsize G.T. Zatsepin,, V.A. Kuz'min,{\it Zh. Eks. Teor.Fiz., Pis'ma Red.{\bf 4} (1966)144}}
%%%%%%\bibitem{zas05}{\normalsize  Zas E. New J.Phys. 7 (2005) 130}
%%%%%\bibitem{Yamamoto2007}{\normalsize T.Yamamoto,Pierre Auger Collaboration,30th ICRC, arXiv:0707.2638,v3 }



\
\end{thebibliography}
\end{document}